\documentclass[letterpaper,12pt]{article}   
\usepackage{osajnl2} 
\usepackage[draft]{hyperref} 
\usepackage{url}

\begin{document}

\title{Polarization and far-field diffraction patterns of total internal
reflection corner cubes}

\author{T.\,W. Murphy, Jr.$^{1,*}$ and S.\,D. Goodrow$^{1}$}
\address{$^{1}$Center for Astrophysics and Space Sciences, University of California, San Diego, \\ 9500 Gilman Drive, MC-0424, La Jolla, CA 92093-0424, USA}
\address{$^{*}$Corresponding author: \tt{tmurphy@physics.ucsd.edu}}

\begin{abstract}

Many corner cube prisms, or retroreflectors, employ total internal reflection (TIR) via uncoated rear surfaces.  The different elliptical polarization states emerging from the six unique paths through the corner cube complicate the far-field diffraction pattern by introducing various phase delays between the six paths. In this paper, we present a computational framework to evaluate polarization through TIR corner cubes for arbitrary incidence angles and input polarization states, presenting example output for key normal-incidence conditions. We also describe a method to produce far-field diffraction patterns resulting from the polarization analysis, presenting representative images---broken into orthogonal polarizations, and characterizing key cases. Laboratory confirmation is also presented for both polarization states and far-field diffraction patterns.

\end{abstract}


\maketitle

\section{Introduction}

Solid glass corner cube prisms (or, more generally, corner cube retroreflectors
or CCRs) are used in interferometers, surveying references, gravimeters,
and for laser ranging to satellites and the Moon. CCRs may either
have a metallic reflective coating on the rear surface, such as silver
or aluminum, or be uncoated to operate via total internal reflection
(TIR). Within $17^{\circ}$ of normal-incidence, TIR CCRs reflect
100\% of the incident light at any azimuthal angle, ignoring reflection
losses at the front surface (which may be anti-reflection coated).
Comparatively, silver coatings operating at 96\% will lose 12\% of
the flux due to three rear surface reflections, and aluminum coatings at 91\% will
sacrifice 25\% of the light. For some applications, absorption of
incident light (e.g., sunlight) by the reflective coating results
in strong thermal gradients within the prism, in turn leading to phase
distortions that disturb the far-field diffraction pattern. In these
cases, TIR cubes are preferred.

On the other hand, coated corner cubes have little effect on the input
polarization state, so that in the absence of thermal gradients or other
distorting influences, the far-field diffraction pattern from such a corner
cube will approach that of a perfect Airy pattern corresponding to the
circular aperture of the corner cube. TIR corner cubes, by contrast,
generally introduce elliptical polarization at each reflection. Each of the
six surface sequence permutations will in general produce a different
output polarization, corresponding to phase offsets between the six paths.
The resulting far-field diffraction pattern for a fused silica CCR has a
central intensity only 26\% that of the perfect reflector case.  Only
36.1\% of the total flux falls within a radius of
$1.22\lambda/D$---corresponding to the first null in the Airy
pattern---where $\lambda$ is the wavelength and $D$ is the diameter of the
corner cube aperture.  The comparable measure for the Airy function is
83.8\%.

The literature contains a number of papers describing polarization and
diffraction of TIR CCRs, but some are inconsistent with each other, and
none of them provide an adequate framework for a comprehensive assessment
of CCR performance compatible with our goals. Specifically, Peck (1962)
\cite{peck} finds polarization eigenmodes for TIR CCRs at normal
incidence---primarily with an interest in using CCRs in optical cavities.
Liu and Azzam (1995) \cite{liu-azzam} offer a comprehensive treatment of
the polarization states emerging from TIR CCRs, along with laboratory
measurements of Stokes parameters. The focus follows that of Peck:
calculating eigenmodes in a coordinate system that has a reflection
relative to the input coordinates. Hodgson and Chipman (1990)
\cite{hodgson} also present laboratory data along with a mathematical
development, but we find the results to be incompatible with ours and other
works---as if the solid cube under examination employed reflective coatings
rather than TIR.  Scholl (1995) \cite{scholl} performs raytrace analysis to
track the state of the electric field within imperfect corner cubes, but
does not treat TIR explicitly. Chang et al. (1971) \cite{chang} provide an
impressive analytic calculation of the far-field diffraction pattern of a
TIR CCR at normal incidence and linear input polarization, along with some
useful quantitative handles. This paper also separates the diffraction
patterns into orthogonal polarization states and provides laboratory checks
on the results, which prompted us to use this paper as a useful standard
against which to compare our normal-incidence linear polarization results.
In a related vein, Arnold produced a series of special reports on methods
for calculating CCR transfer functions \cite{arnold}.  Most recently,
Sadovnikov and Sokolov (2009) \cite{sadovnikov}, and later Sokolov and
Murashkin (2011) \cite{sokolov}, contribute a work most similar to our own,
presenting diagrams of polarization and diffraction patterns at different
input polarizations for the normal incidence case.  However, the works were
not readily adaptable to our needs because: 1) coordinate systems and
plotting conventions are not clearly described; 2) the corner cubes
considered do not appear to be circularly cut; and 3) the presentation is
not geared toward instructing readers on how to develop their own analysis
capability---as this work aims to do.

Our ultimate goal is to assess the far-field diffraction pattern from TIR
CCRs subject to thermal gradients for application in our lunar laser
ranging project \cite{apollo} (see the companion paper on thermal gradients
within CCRs \cite{ccr-thermal}). Because the target CCR is in relative
tangential motion with respect to the line of sight, velocity aberration
shifts the pattern relative to our receiving telescope.  We therefore
sample the shoulder of the central diffraction peak, and thus are not
content with knowledge of the central irradiance of the diffraction
pattern. Even though the lunar CCRs are designed to minimize thermal
gradients, we observe strong evidence that thermal gradients are developing
at certain lunar phases---likely due to solar illumination of dust
deposited on the front faces of the prisms \cite{dust}. We have found the
existing literature to be insufficient for prescribing analysis algorithms that
we might emulate, and further found inadequate 
published experimental results against which to verify our results.

We describe here a technique to analyze corner cube polarization and
diffraction patterns at arbitrary angles of incidence that should be
straightforward to program into a computer language (we used Python, and
make our code available online).  Moreover, we display graphical output of
polarization states and of diffraction patterns that should be useful for
comparison and as a demonstration of the general behavior of TIR CCR
diffraction. Laboratory polarization measurements confirm the analysis, and
far-field diffraction patterns verify the final result.

\section{Corner Cube Geometry and Raytracing\label{sec:geometry}}

\begin{figure}
\begin{center}\includegraphics[%
  scale=0.6]{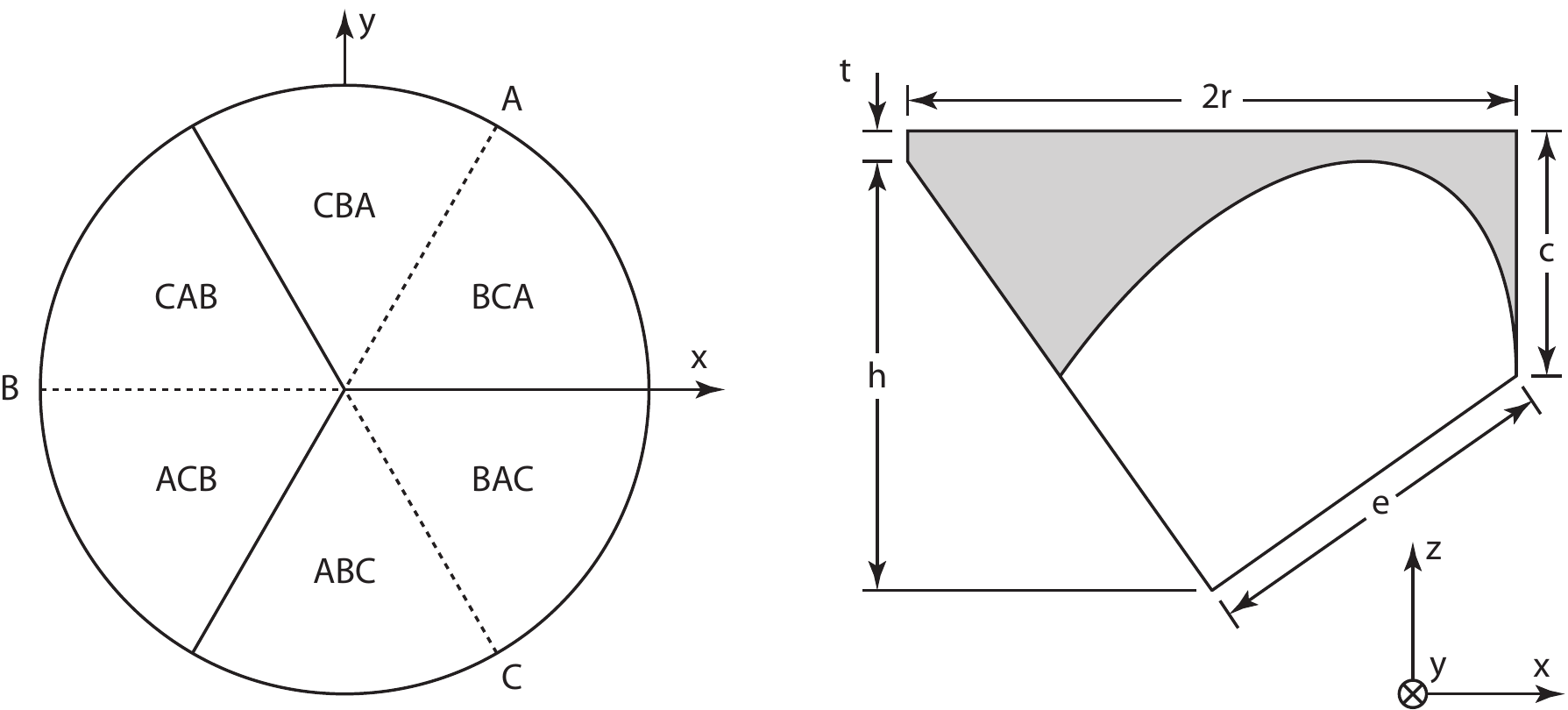}\end{center}

\caption{Corner cube geometry and global coordinate system. The three back
faces are labeled A, B, and C. Dotted lines represent reflections
of the real edges. Three-letter sequences placed on each ``wedge''
identify the exit location for the six unique paths through the CCR
(the corresponding input wedge is diametrically opposite). The view
at right is along the global $y$-axis, with face C exposed to view. In units of
the circular radius, $r$, $h=\sqrt{2}$, $e=\sqrt{\frac{3}{2}}$,
and $c=\sqrt{\frac{1}{2}}$. The distance $t$ is arbitrary, representing
the height of the uninterrupted cylinder around the CCR.\label{fig:CCR-geom}}
\end{figure}

Figure~\ref{fig:CCR-geom} depicts the geometry and orientation of
a circularly cut CCR within a global right-handed Cartesian coordinate
system. Rear faces are labeled A, B, and C, so that a particular ray
path through the CCR can be labeled as ACB, for instance. 

The three normal vectors for the rear surfaces form an orthonormal
set:\begin{equation}
\begin{array}{c}
\hat{\mathbf{n}}_{A}=\frac{1}{\sqrt{6}}\left(\begin{array}{c}
-1\\
-\sqrt{3}\\
\sqrt{2}\end{array}\right)\\
\hat{\mathbf{n}}_{B}=\frac{1}{\sqrt{6}}\left(\begin{array}{c}
2\\
0\\
\sqrt{2}\end{array}\right)\\
\hat{\mathbf{n}}_{C}=\frac{1}{\sqrt{6}}\left(\begin{array}{c}
-1\\
\sqrt{3}\\
\sqrt{2}\end{array}\right)\end{array}.\label{eq:normals}\end{equation}
We define the distant observer's angular position relative to the
CCR by an azimuth, $A$---measured from the $x$--axis and increasing
toward the $y$--axis---and an inclination, $i$---away from the $z$--axis,
so that $\hat{\mathbf{k}}_{0}=\left< -\sin i\cos A,\,-\sin i\sin A,\,-\cos i\right>$.
Snell's law can be applied at the CCR front face to redirect an incident
light ray into a new $\hat{\mathbf{k}}$, while reflection within
the CCR changes the ray direction according to $\hat{\mathbf{k}}\rightarrow\hat{\mathbf{k}}-2(\hat{\mathbf{k}}\cdot\hat{\mathbf{n}})\hat{\mathbf{n}}$,
where $\hat{\mathbf{n}}$ is the surface normal in question. 

We define a frame for input polarization that we associate with horizontal
($\hat{\mathbf{s}}_{0}$) and
vertical ($\hat{\mathbf{p}}_{0}$) in such a way that the horizontal unit vector is perpendicular
to both $\hat{\mathbf{k}}_{0}$ and $\hat{\mathbf{z}}$, which itself
is the front surface normal. Explicitly, \begin{equation}
\begin{array}{c}
\hat{\mathbf{s}}_{0}=\left< -\sin A,\,\cos A,\,0\right>\\
\hat{\mathbf{p}}_{0}=\hat{\mathbf{s}}_{0}\times\hat{\mathbf{k}}_{0}\end{array},\label{eq:s0-p0}\end{equation}
where $A$, again, is the azimuth of the observer.  We will
present both input and output polarization states in the
globally-referenced observer frame of
Eq.~\ref{eq:s0-p0}, which will ultimately require a coordinate flip owing to the
retroreflection.

\begin{figure}
\begin{center}\includegraphics[%
  scale=0.75]{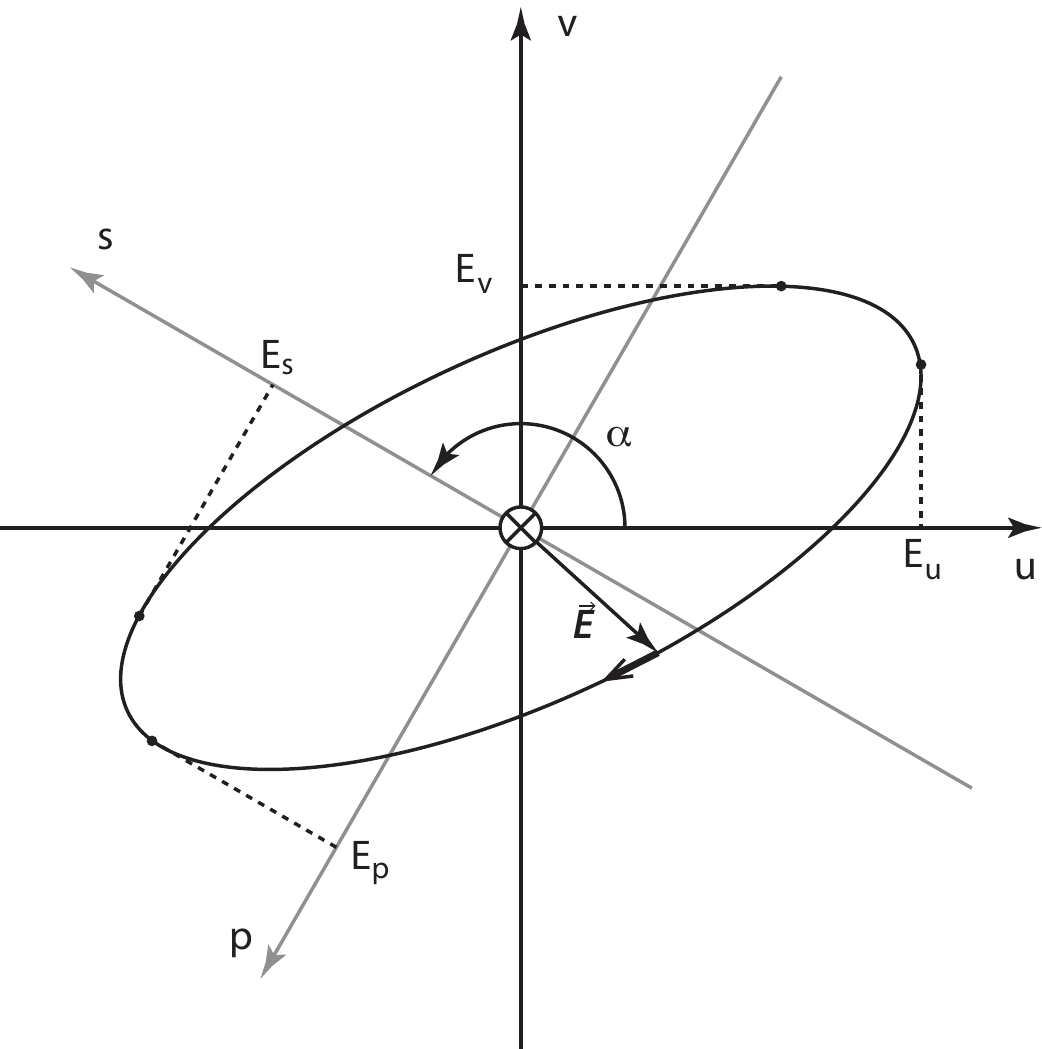}\end{center}

\caption{Coordinate system for representation of elliptical polarization,
looking along the propagation direction (note central $\times$ denoting 
$\hat{\mathbf{k}}$ going into the page). By the conventional definition,
the electric field vector pictured rotates in a left-handed sense, when looking
toward the light source.\label{fig:coord-ellipse}} \end{figure}

On approach to each interface one must
transform into the local $s$ and $p$ coordinate system corresponding
to directions perpendicular and parallel to the plane of incidence,
respectively. The $s$--$p$ frame is described by\begin{equation}
\begin{array}{c}
\hat{\mathbf{s}}=\frac{\hat{\mathbf{k}}\times\hat{\mathbf{n}}}{\left|\hat{\mathbf{k}}\times\hat{\mathbf{n}}\right|}\\
\hat{\mathbf{p}}=\hat{\mathbf{s}}\times\hat{\mathbf{k}}\end{array},\label{eq:s-p-def}\end{equation}
which happens to be aligned with the global $x$--$y$ frame for light approaching the corner cube from azimuth $A=-90^\circ$, and appearing right-handed if looking along $\hat{\mathbf{k}}$.
The transformation between some arbitrary
$u$--$v$ frame perpendicular to the propagation direction and the
$s$--$p$ frame for the upcoming surface interface can be determined
from the four-quadrant arctangent\begin{equation}
\alpha=\mathrm{atan2}(\hat{\mathbf{s}}\cdot\hat{\mathbf{v}},\hat{\mathbf{s}}\cdot\hat{\mathbf{u}}),\label{eq:alpha-det}\end{equation}
as depicted in Figure~\ref{fig:coord-ellipse}. After the interface---whether
refractive or reflective---the propagation direction, $\hat{\mathbf{k}}$,
is altered by some rotation about the $\hat{\mathbf{s}}$ direction.
Consequently, $\hat{\mathbf{s}}$ is unchanged at a single interface, while $\hat{\mathbf{p}}$
must be re-evaluated according to Eq.~\ref{eq:s-p-def}. As one steps
through the corner cube, the $\hat{\mathbf{s}}$ and $\hat{\mathbf{p}}$
vectors become the $\hat{\mathbf{u}}$ and $\hat{\mathbf{v}}$ vectors
for the next application of Eq.~\ref{eq:alpha-det}.

For reference, the rotation angles, $\alpha$, for all six path sequences
through the corner cube at normal incidence are given in
Table~\ref{tab:rotations}, where the initial $u$--$v$ coordinate system is
aligned to the global $x$--$y$ frame ($A=-90^{\circ}$). The last rotation,
$\alpha_{4}$, aligns the final $\hat{\mathbf{p}}$ vector with
$\hat{\mathbf{p}}_0$, while the retroreflection
($\hat{\mathbf{k}}\rightarrow -\hat{\mathbf{k}}_0$) results in a coordinate
flip so that $\hat{\mathbf{s}}$ points along $-\hat{\mathbf{s}}_0$.

\begin{table}

\caption{Rotation Sequences, in Degrees\label{tab:rotations}}

\begin{center}\begin{tabular}{lrrrr}
\hline 
path&
$\alpha_{1}$&
$\alpha_{2}$&
$\alpha_{3}$&
$\alpha_{4}$\\
\hline 
ACB&
$150$&
$-60$&
$60$&
$-90$\\
ABC&
$150$&
$60$&
$-60$&
$30$\\
BAC&
$-90$&
$-60$&
$60$&
$30$\\
BCA&
$-90$&
$60$&
$-60$&
$150$\\
CBA&
$30$&
$-60$&
$60$&
$150$\\
CAB&
$30$&
$60$&
$-60$&
$-90$\\
\hline
\end{tabular}\end{center}
\end{table}

\section{Polarization and Phases}

We describe the electric field transverse to the direction of propagation
by a two-component vector in an orthogonal basis, expressed in the
global 3-D coordinate system by the unit vectors $\hat{\mathbf{u}}$
and $\hat{\mathbf{v}}$. Because pathlengths within a perfect CCR
are independent of position for a given $\hat{\mathbf{k}}_{0}$, we
can suppress the phase advance associated with forward propagation,
concentrating only on the temporal and static phase offsets---the
latter changing only at interfaces. In the generalized coordinates,
$u$ and $v$, the electric field vector follows\begin{equation}
\vec{E}=\left(\begin{array}{c}
E_{u}\cos(\omega t+\delta_{u})\\
E_{v}\cos(\omega t+\delta_{v})\end{array}\right),\label{eq:eu-ev}\end{equation}
where $E_{u}$ and $E_{v}$ are positive electric field amplitudes
in the $u$ and $v$ directions and $\delta_{u}$ and $\delta_{v}$
are the associated phases.  The $\omega t$ term represents time evolution of the phase at frequency $\omega$.  For convenience, we normalize the intensity, setting $E_{u}^{2}+E_{v}^{2}=1$.
It is important to track individual phases
rather than just the phase difference between components; while the difference
is sufficient to describe the polarization state, the absolute phases
are important for constructing a far-field diffraction pattern. 

In order to determine the sense of rotation, we can look at the phase
difference, $\delta\equiv\delta_{v}-\delta_{u}$. If we confine $\delta$ to
the range $-\pi<\delta\le\pi$ by adding or subtracting integer multiples of
$2\pi$, we can associate $\delta<0$ with right-hand polarization, and
$\delta>0$ with left-hand polarization, when adopting the convention of
looking toward the light source.  For instance, in
Figure~\ref{fig:coord-ellipse} the electric field vector will reach its
maximum positive value in $v$ ($E_{v}$) shortly before it reaches $E_{u}$.
Therefore $\delta_{u}$ must be slightly less than $\delta_{v}$ in
accordance with Eq.~\ref{eq:eu-ev}, so that $\delta>0$ and we get
left-handed polarization, as depicted. Linear polarization is described by
$\delta=0$ or $\delta=\pi$.

In order to transform the properties of the ellipse through a rotation
by angle $\alpha$, as defined in Eq.~\ref{eq:alpha-det} and depicted
in Figure~\ref{fig:coord-ellipse}, we rotate an arbitrary electric
field vector positioned somewhere on the ellipse by \begin{equation}
\left(\begin{array}{c}
E_{s}\cos(\omega t+\delta_{s})\\
E_{p}\cos(\omega t+\delta_{p})\end{array}\right)=\left(\begin{array}{cc}
\cos\alpha & \sin\alpha\\
-\sin\alpha & \cos\alpha\end{array}\right)\left(\begin{array}{c}
E_{u}\cos(\omega t+\delta_{u})\\
E_{v}\cos(\omega t+\delta_{v})\end{array}\right).\label{eq:uv-to-sp}\end{equation}
We then separately equate all $\cos\omega t$ and $\sin\omega t$
terms in a trigonometric expansion of the terms above to find that
\begin{equation}
\begin{array}{c}
E_{s}\cos\delta_{s}=E_{u}\cos\delta_{u}\cos\alpha+E_{v}\cos\delta_{v}\sin\alpha\\
E_{s}\sin\delta_{s}=E_{u}\sin\delta_{u}\cos\alpha+E_{v}\sin\delta_{v}\sin\alpha\\
E_{p}\cos\delta_{p}=E_{v}\cos\delta_{v}\cos\alpha-E_{u}\cos\delta_{u}\sin\alpha\\
E_{p}\sin\delta_{p}=E_{v}\sin\delta_{v}\cos\alpha-E_{u}\sin\delta_{u}\sin\alpha\end{array}.\label{eq:sp-params}\end{equation}
From this, one may compute the new phases via\begin{equation}
\begin{array}{c}
\delta_{s}=\mathrm{atan2}(E_{s}\sin\delta_{s},\, E_{s}\cos\delta_{s})\\
\delta_{p}=\mathrm{atan2}(E_{p}\sin\delta_{p},\, E_{p}\cos\delta_{p})\end{array},\label{eq:deltas}\end{equation}
which is insensitive to the values of $E_{s}$ and $E_{p}$ because these
factors are common to the numerator and denominator of the arctangent
argument. $E_{s}$ and $E_{p}$ can then be extracted by combining the results for
$\delta_{s}$ and $\delta_{p}$ with Eq.~\ref{eq:sp-params}.  One may verify
that $E_{s}^{2}+E_{p}^{2}=E_{u}^{2}+E_{v}^{2}$ as a check on the
computation.

At the front surface refractive interface, we diminish $E_{s}$ and
$E_{p}$ according to the Fresnel equations---about 3.5\% for fused
silica at normal incidence---with $s$-polarization reflection increasing
for larger incidence angles while $p$-polarization reflection decreases.
Anti-reflection coatings would modify this procedure.

At each reflective interface within the CCR, the values of $E_{s}$
and $E_{p}$ will be preserved---either in TIR or for a perfect reflector.
The phases, however, will shift according to the Fresnel relations
for TIR:
\begin{equation}
\begin{array}{c}
\delta_{s}\rightarrow\delta_{s}+\Delta\delta_{s}\\
\delta_{p}\rightarrow\delta_{p}+\Delta\delta_{p}\end{array},\label{eq:update}\end{equation}
 where\begin{equation}
\begin{array}{c}
\Delta\delta_{s}=2\tan^{-1}\left(\frac{\sqrt{n^{2}\sin^{2}\theta-1}}{n\cos\theta}\right)\\
\Delta\delta_{p}=2\tan^{-1}\left(\frac{n\sqrt{n^{2}\sin^{2}\theta-1}}{\cos\theta}\right)\end{array},\label{eq:fresnel-shift}\end{equation}
with $n$ being the refractive index of the medium (assuming vacuum on the
other side) and $\theta$ being the angle of incidence determined by
$\cos\theta=\left|\hat{\mathbf{k}}\cdot\hat{\mathbf{n}}\right|$.  At normal
incidence, each reflection has $\cos\theta=\frac{1}{\sqrt{3}}$ ($\theta
\approx54.74^{\circ})$, so that fused silica at $n\approx1.46$ results in
$\Delta\delta_{s}\approx 1.31$ rad, and $\Delta\delta_{p}\approx 2.05$~rad.
Our choice of conventions (e.g., Eq.~\ref{eq:eu-ev}) demands positive signs
for Eq.~\ref{eq:fresnel-shift} to match experimental results both in terms
of polarization ellipses and diffraction pattern orientations.  For testing
purposes, it is often useful to model perfect reflection, in which case
$\delta_{p}$ is unchanged, while $\delta_{s}$ changes by $\pi$ at each
interface.  In this case, the orientation and elliptical aspect of any
polarization state is preserved in the global coordinate system on
completing passage through the CCR, while the rotational sense switches
handedness.

We therefore have a complete description of the procedure for tracking
the four polarization parameters through the corner cube. On approach
to each surface, the rotation angle of the current coordinate frame relative
to the upcoming $s$--$p$ frame is found; the polarization parameters
are rotated into this frame; the phases are updated; the outbound
$\hat{\mathbf{k}}$ and $\hat{\mathbf{p}}$ vectors are established;
and the procedure repeats. Example code that can replicate all the
results in this paper can be found at \url{http://physics.ucsd.edu/~tmurphy/papers/ccr-sim/ccr-sim.html}.

\subsection{Matrix Approach}

For normal incidence, we can cast the procedure into a Jones matrix approach:
\begin{equation}
\mathbf{T}=\mathbf{F}\cdot\mathbf{R}(\alpha_{4})\cdot\mathbf{P}\cdot\mathbf{R}(\alpha_{3})\cdot\mathbf{P}\cdot\mathbf{R}(\alpha_{2})\cdot\mathbf{P}\cdot\mathbf{R}(\alpha_{1}),\label{eq:trans-matrix}\end{equation}
where the rotation matrices, $\mathbf{R}$, use the angles provided
in Table~\ref{tab:rotations} according to
\begin{equation}
\mathbf{R}(\alpha)=\left(\begin{array}{cc}
\cos\alpha & \sin\alpha\\
-\sin\alpha & \cos\alpha\end{array}\right).\label{eq:rotmatrix}\end{equation}
The Jones matrix, $\mathbf{P}$, is a diagonal matrix for advancing
the phase of $\delta_{s}$ and $\delta_{p}$:
\begin{equation}
\mathbf{P}=\left(\begin{array}{cc}
e^{i\Delta\delta_{s}} & 0\\
0 & e^{i\Delta\delta_{p}}\end{array}\right),\label{eq:phase-matrix}\end{equation}
where the phase shifts are given by Eq.~\ref{eq:fresnel-shift}.
Finally, in keeping with our approach in this paper of representing
output states in the global coordinate frame, while the propagation
direction has turned $180^{\circ}$, we apply a coordinate reflection,
\begin{equation}
\mathbf{F}=\left(\begin{array}{cc}
-1 & 0\\
0 & 1\end{array}\right),\label{eq:flip-matrix}\end{equation}
to the result.  We have left out the reflection loss from the front surface
to simplify the presentation.

For example, the composite matrix for the ACB path
is\begin{equation}
\mathbf{T}_{\mathrm{ACB}}=\left(\begin{array}{cc}
0.655e^{2.78i} & 0.755e^{2.24i}\\
0.755e^{1.51i} & 0.655e^{-2.16i}\end{array}\right).\label{eq:example-matrix}\end{equation}
We apply this matrix to an input polarization vector similar to that
in Eq.~\ref{eq:eu-ev}. For instance we can describe a linear polarization
input by the vector $\mathbf{P}_{\mathrm{in}}=\left< \cos\theta,\ \sin\theta\right>$,
where $\theta=0$ represents polarization along the global $x$-axis.
Circular polarization would have the vector $\mathbf{P}_{\mathrm{in}}=\left<1,\ \pm i\right> /\sqrt{2}$.
We then form the output polarization vector: $\mathbf{P}_{\mathrm{out}}=\mathbf{T}\mathbf{P}_{\mathrm{in}}$.
For example, if $\theta=45^{\circ}$, we find that the ACB path produces
$\mathbf{P}_{\mathrm{out}}=\left< 0.962e^{2.49i},\ 0.272e^{2.56i}\right>$. Given
the amplitudes $P_{x}$ and $P_{y}$ and the phase difference $\delta=\delta_{y}-\delta_{x}\approx 0.07$
in this case, we can find the polarization ellipse parameters by first
constructing
\begin{equation}
\tan2\omega t=-\frac{P_{y}^{2}\sin2\delta}{P_{x}^{2}+P_{y}^{2}\cos2\delta},
\label{eq:tan2wt}\end{equation}
solving for $\omega t$, then producing the ellipse vertices by:
\begin{equation}
\begin{array}{rcl}
x_{1} & = & P_{x}\cos\omega t\\
y_{1} & = & P_{y}\cos(\omega t+\delta)\\
x_{2} & = & P_{x}\cos(\omega t+\frac{\pi}{2})\\
y_{2} & = & P_{y}\cos(\omega t+\frac{\pi}{2}+\delta)\end{array},
\label{eq:ellipse-extrema}\end{equation}
after which one computes the semi-major and semi-minor axes via the
Pythagorean distances from the origin created by coordinate pairs
$(x_{1},\  y_{1})$ and $(x_{2},\  y_{2})$. The angle from the $x$-axis
is then calculated as $\psi=\arctan(y/x)$ for the coordinate pair
associated with the major axis. In the example case of $45^{\circ}$
linear polarization following path ACB, we find that $a=0.9998$,
$b=0.019$, and $\psi=15.8^{\circ}$. The state is nearly linear,
and can be picked out in the fourth panel of Figure~\ref{fig:lin-pol-out}.

One must take care in interpreting the rotational sense of
$\mathbf{P}_{\mathrm{out}}$, because the
coordinate flip matrix, $\mathbf{F}$, amounts to a reversal of
$\hat{\mathbf{k}}$ relative to the $s$--$p$ frame, reversing the
association between the sign of $\delta$ and handedness.  In this example
case, with $\delta\approx0.07$, the state is right-handed.  Presenting
polarization states in a global frame when $\hat{\mathbf{k}}$ turns
$180^\circ$ inevitably invites complication of this sort.

\section{Polarization Results}

\begin{figure}
\begin{center}
\hfill{}\includegraphics[%
  height=1.25in,
  keepaspectratio]{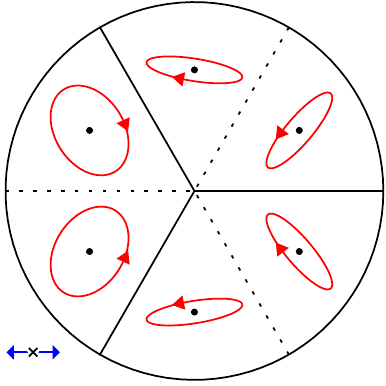}\hfill{}\includegraphics[%
  height=1.25in,
  keepaspectratio]{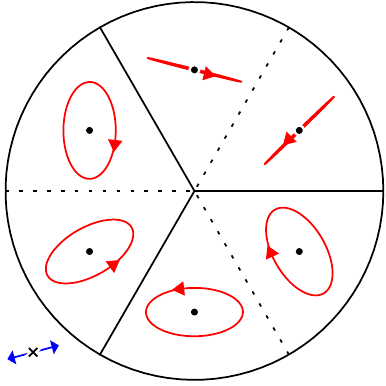}\hfill{}\includegraphics[%
  height=1.25in,
  keepaspectratio]{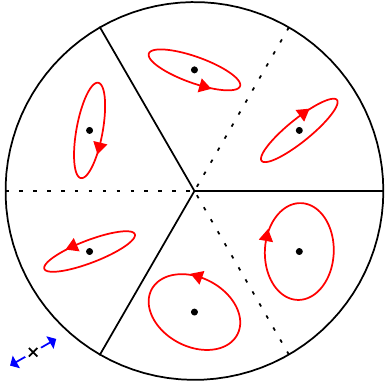}\hfill{}\includegraphics[%
  height=1.25in,
  keepaspectratio]{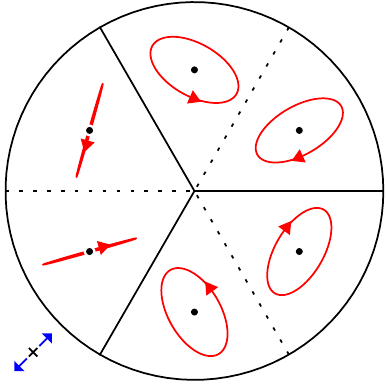}\hfill{}\includegraphics[%
  height=1.25in,
  keepaspectratio]{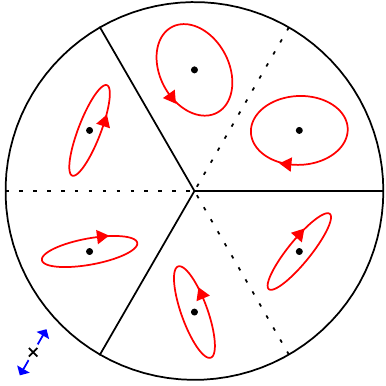}\hfill{}
\end{center}

\caption{Output polarization states at normal incidence for linear input
polarization rotated in $15^{\circ}$ increments. The input polarization is
depicted at lower left, with a cross to indicate light traveling into the page.
The output polarization is drawn in the wedge from which it emerges, as it
would be oriented in the frame looking at the CCR face. Light output
emerges from the page in this view (indicated by the dot in the center of
each ellipse),
so that right-handed polarization states show clockwise rotation.  The
pattern in the rightmost frame matches that in the leftmost frame with a
120$^{\circ}$ rotation.\label{fig:lin-pol-out}} \end{figure}

Figure~\ref{fig:lin-pol-out} shows the output polarization states
computed for a fused silica CCR at normal incidence with a refractive
index of around 1.46. After $60^{\circ}$ degrees of rotation, the
pattern repeats, albeit with an additional $180^{\circ}$ rotation. Therefore,
a $120^{\circ}$ rotation results in an exact replication of the pattern
with respect to the corner cube,
in accord with the three-fold symmetry of the CCR. Tracking the output
of a particular wedge reveals a smooth stepwise progression through
ellipse eccentricity and rotation sense. Within each wedge, the orientation of the major
axis tends to rotate slowly in a direction counter to the stepwise evolution
of the input polarization angle.

\begin{figure}
\hfill{}\includegraphics{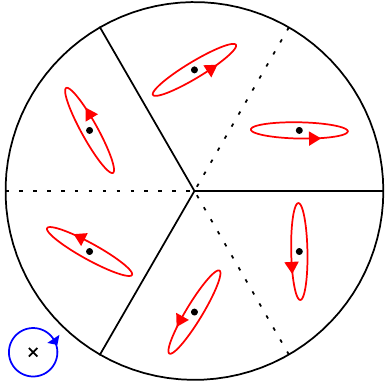}\hfill{}\includegraphics{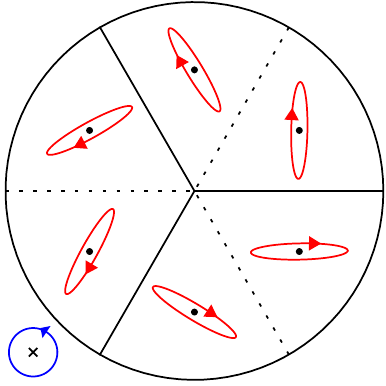}\hfill{}

\caption{Output polarization states at normal incidence for circular input
polarization. Parameters and conventions are as described for Figure~\ref{fig:lin-pol-out}.
The input light is depicted traveling into the page, so that left-handed
polarization is seen at left, and right-handed at right.\label{fig:circ-pol-out}}
\end{figure}

For circular input polarization at normal incidence, there is no need
to explore orientation changes. Figure~\ref{fig:circ-pol-out} shows
the rather symmetric polarization output patterns given circular input
polarization into fused silica. For fused silica, the minor-to-major
axis ratio is 0.168, while for BK7 it is 0.121. The output ellipses
emerge with the same polarization sense as the input, although the
arrows in the figure appear to be reversed on account of the reversal
of light propagation direction.

As a computational check, Table~\ref{tab:amp-phase-ex} provides
a sample of amplitudes and phases in the global $x$--$y$ frame for
normal incidence light polarized along the $x$-direction, leaving out the
front-surface reflection loss.  This corresponds
to the left-most panel in Figure~\ref{fig:lin-pol-out}. We use a
refractive index of 1.45702, corresponding to fused silica at 632.8~nm.
Lurking in the $\delta_{y}$ column are phase pairs differing by $\pi$.
Being at normal incidence, these results can be reproduced via the
matrix method of Eq.~\ref{eq:trans-matrix}, and the first row of
the table is represented by the example matrix in Eq.~\ref{eq:example-matrix}.

\begin{table}

\caption{Example Electric Field Parameters$^a$ \label{tab:amp-phase-ex}}

\begin{center}\begin{tabular}{lcccc}
\hline 
Path&
$E_{x}$&
$\delta_{x}$&
$E_{y}$&
$\delta_{y}$\\
\hline 
ACB&
$0.65547$&
$2.77848$&
$0.75523$&
$1.51218$\\
ABC&
$0.96282$&
$-1.82634$&
$0.27014$&
$-2.83442$\\
BAC&
$0.65547$&
$2.77848$&
$0.75523$&
$-0.89783$\\
BCA&
$0.65547$&
$2.77848$&
$0.75523$&
$2.24376$\\
CBA&
$0.96282$&
$-1.82634$&
$0.27014$&
$0.30718$\\
CAB&
$0.65547$&
$2.77848$&
$0.75523$&
$-1.62941$\\
\hline
\end{tabular}
\end{center}
\footnotesize $^a$For normal incidence, horizontal input polarization
\normalsize
\end{table}

\subsection{Experimental Comparison}

\begin{figure}
\hfill{}\includegraphics[%
  height=1.5in,
  keepaspectratio]{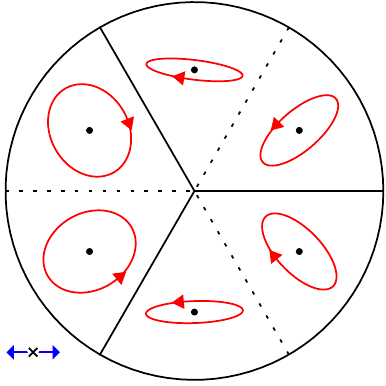}\hfill{}\includegraphics[%
  height=1.5in,
  keepaspectratio]{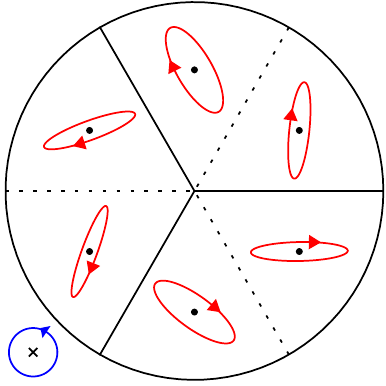}\hfill{}

\caption{Experimental polarization results, plotted following conventions
in Figures~\ref{fig:lin-pol-out} and \ref{fig:circ-pol-out}. At
left is linear polarization matching the left-most panel in Figure~\ref{fig:lin-pol-out},
and at right is right-handed polarization input. Slight irregularities
are discussed in the text, but the overall agreement with theoretical
expectations is good.\label{fig:pol-meas}}
\end{figure}

Using a fused silica CCR and a HeNe laser at 632.8~nm, we directed
a high-purity ($10^{5}$:1 intensity ratio) linear polarization state into each path
sequence in turn, characterizing the emerging elliptical polarization
state in terms of major and minor axes (taking the square root of
measured intensity to find electric field amplitude), angle of the axis,
and rotation sense with the help of a high-precision quarter-wave
plate. The quarter-wave plate also provided an independent check of
the ellipse axis ratio---this time directly as an electric field ratio.
We employed a separate precision quarter-wave plate to send circular
polarization into the CCR, confirming an axis ratio of 0.99 in amplitude.
Figure~\ref{fig:pol-meas} shows the results for two cases, in a
format similar to that of previous plots. 

We found in practice that the measured ellipse properties deviated more
than we expected, given the purity of input polarization (see, for
instance, the minor axis variations for the circular polarization case in
Figure~\ref{fig:pol-meas}). Anomalies did not follow the CCR upon a
$120^{\circ}$ rotation of the CCR with respect to the optical bench, but
stayed fixed in the laboratory frame, suggesting that the discrepancy
resides in the measurement setup. The orientation of the major axis tends
to be robust (within $10^{\circ}$), as this is a result of gross rotations
(projections) of the input electric field vectors---both of which are
controlled or known to adequate precision. The axis ratio,
however, is very sensitive to phase differences between orthogonal
polarizations, and could vary substantially.  For the circular polarization
case in Figure~\ref{fig:pol-meas}, the minor axis amplitude varies from 0.16 to 0.34
(expecting 0.17), while the corresponding phase differences ($\delta$,
evaluated in a frame where the major axis has $\psi = 45^{\circ}$) remain
within $20^{\circ}$ of theoretical expectations.  For the linear
polarization case, phase differences stayed within $15^{\circ}$ of the
expected values. Given the high degree of fidelity we observe in the
far-field diffraction patterns---as demonstrated below---we conclude that
the polarization states are indeed following the model closely, even if the
results in Figure~\ref{fig:pol-meas} do not appear to be an exact match.

\section{Diffraction Method}

The far-field diffraction pattern can be conveniently calculated via
the Fourier transform (FT) of the complex amplitude and phase of the
electric field at the exit aperture of the corner cube. The FT integrates
area-weighted amplitude and phase contributions at the aperture, resulting
in the net sum---or interference---of the electric field at infinite
distance as a function of angular displacement from the propagation
direction. The square magnitude of the FT then represents the intensity
in the far-field:\begin{equation}
I(\chi,\eta)=\left|\mathrm{\int\int}_{\mathrm{aperture}}S(u,v)\exp\left[i\phi(u,v)\right]\exp\left[ik(\chi u+\eta v)\right]\mathrm{d}u\mathrm{d}v\right|^{2},\label{eq:fourier-transform}\end{equation}
where the aperture amplitude, $S$, and phase, $\phi$, are functions
of coordinates $u$ and $v$ in the aperture plane. The coordinates
$\chi$ and $\eta$ then represent angular coordinates in the far-field,
with $k=2\pi/\lambda$.

Orthogonal polarizations cannot interfere with each other, so the
Fourier transform must be broken into separate computations for any
two orthogonal polarizations. For each, the phases are simply the
final $\delta_{u}$ and $\delta_{v}$ phases resulting from the transformation
of the final values of $\delta_{s}$ and $\delta_{p}$ computed via
sequential applications of Eq.~\ref{eq:fresnel-shift} into some
final $u$--$v$ coordinate frame. Each wedge in the aperture---corresponding
to each of the six unique path sequences---will have constant phase
across the wedge.

The aperture function can be determined during preparation for the
Fourier transform, in that one must pass to the integral a two-dimensional
array of aperture amplitudes in the input $u$--$v$ frame. By raytracing
a grid of input ray positions sharing the same input $\hat{\mathbf{k}}$
vector, one can determine which rays emerge by rejecting any ray that
encounters any of the four CCR planes outside the cylindrical radius
of the CCR. The resulting aperture for non-normal incidence has a
shape given by the included intersection of two equal ellipses shifted relative
to one another along their minor axes, each one representing
the projected rim of the entrance aperture and the retro-reflected
rendition of the same. The raytrace also determines which sequence
(wedge) applies, and thus which amplitudes among the set of six pre-computed
$E_{u}$ and $E_{v}$ values are to be used for $S(u,v)$.

One can readily compute the central irradiance, $I(0,0)$, of the
far-field diffraction pattern in the normal incidence case simply
by summing the aperture function, $S(u,v)$, times the phase function,
$\exp\left[i\phi(u,v)\right]$, equally weighted for all six wedges.
In the trivial case where $S=1$ inside a circular aperture of radius
$R$ while $\phi$ is constant, we get a central irradiance of $\pi^{2}R^{4}$.
Summing the values in Table~\ref{tab:amp-phase-ex} (where $E\rightarrow S$
and $\delta\rightarrow\phi$), each weighted by $\pi R^{2}/6$, we
get a central irradiance for the $x$-component of:
\begin{equation}
I(0,0)=\left|\frac{\pi R^{2}}{6}\sum_{n=1}^{6}S_{n}e^{i\phi_{n}}\right|^{2}\approx0.264\pi^{2}R^{4},\label{eq:sector-sum}
\end{equation}
and a $y$-component summing to zero. Thus the total central irradiance
of the TIR diffraction pattern is 26.4\% of what it would be for a
perfect Airy pattern. Combining this with reflection loss from an
uncoated fused silica front surface (incurred twice) puts the central irradiance
at 24.6\% that of the Airy pattern for a circular aperture of the same diameter.

We can develop a useful tool for computing the expected central irradiance
in the far-field diffraction pattern if we characterize all the flux as
being contained in a tophat pattern whose uniform intensity is set to that of
the central peak of the actual diffraction pattern.  This crude model
permits a simple estimation of the central intensity, once the tophat
diameter is known.  Expressed in terms of the diffraction scale, an
uncoated fused silica CCR is characterized by a tophat diameter of
2.56$\lambda/D$.  For the Apollo corner cubes at 532~nm, this is
7.4~arcsec. The corresponding measure for a perfect Airy pattern is
1.27$\lambda/D$. Conversely, if we conveniently---albeit
na\"ively---modeled the diffraction pattern as containing all the flux
within a tophat diameter set to $\lambda/D$, we find that the central
irradiance of the actual pattern is reduced to 0.152 times the nominal
value suggested by the simple $\lambda/D$ tophat model.  Reflection losses
at the front surface degrade the performance further.

\section{Far-field Diffraction Results}

In the diffraction patterns we present, the orientation convention is in
keeping with those in the rest of the paper: looking at the corner cube.
Thus the global $+\hat{\mathbf{x}}$ direction is to the right, and
$+\hat{\mathbf{y}}$ is up.  Direction cosines are plotted, so that light
arriving at positive-$x$ global coordinates in the far field are shown to
the right.  If projected onto a screen at infinity, each of the images here
would incur a left-right flip.  The horizontal direction follows that used
to define polarization, being perpendicular to the plane of incidence and
therefore lying in the global $x$--$y$ plane.

\begin{figure}
\begin{center}\includegraphics[%
  height=2.5in,
  keepaspectratio]{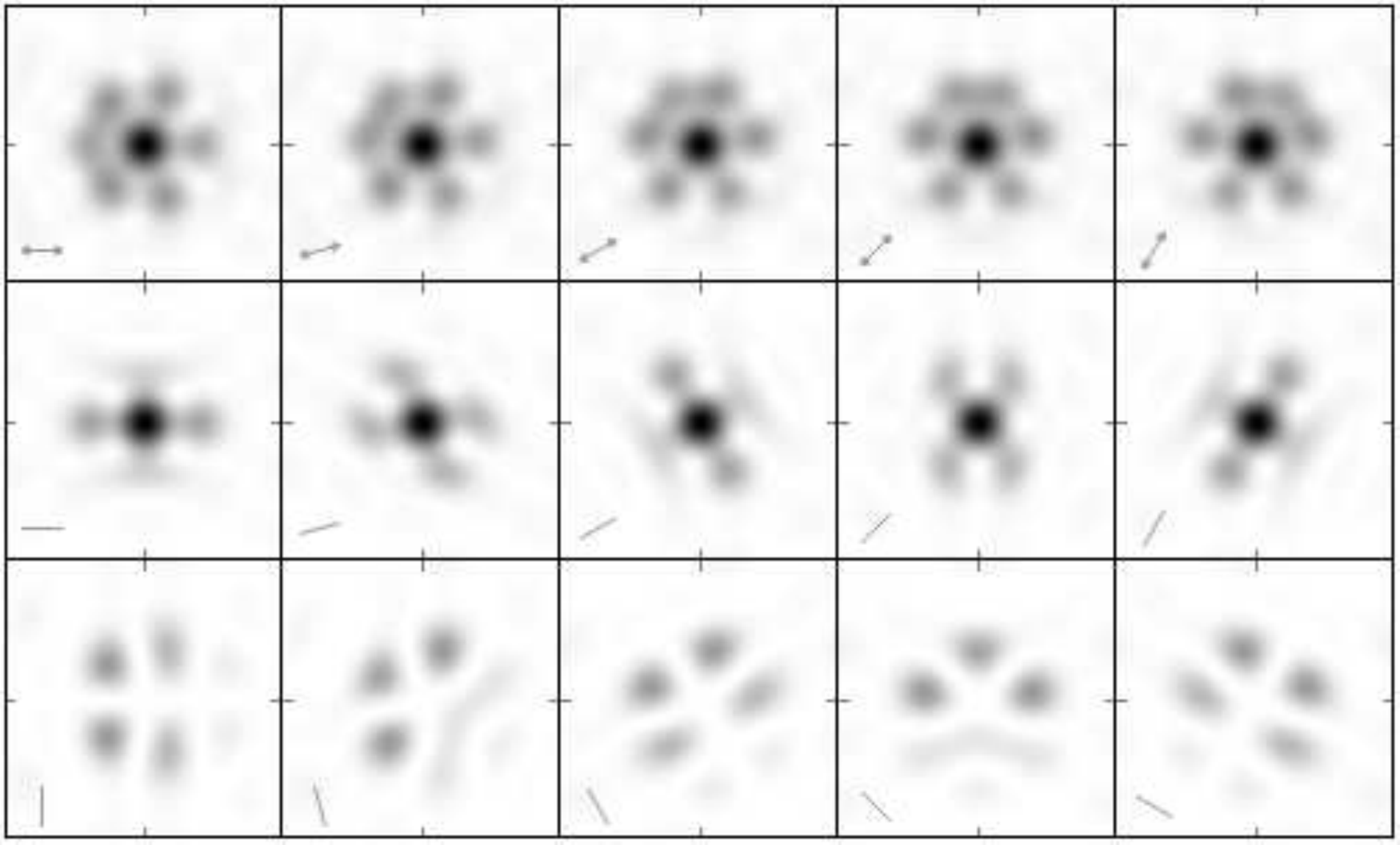}\hspace{20mm}\includegraphics[%
  height=2.5in,
  keepaspectratio]{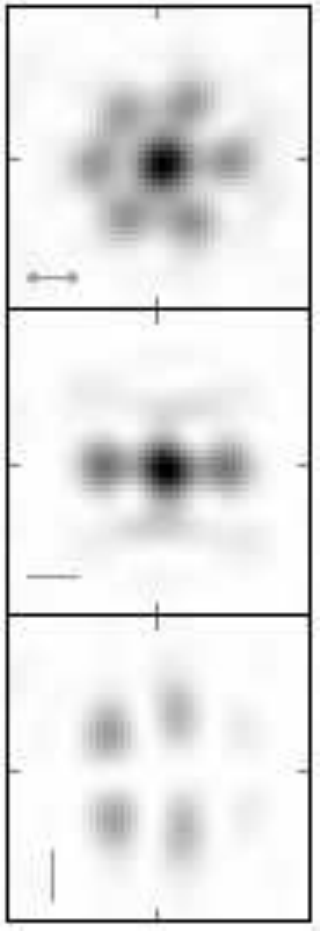}\end{center}

\caption{Normal incidence far-field diffraction patterns for five orientations
of linear polarization input in $15^{\circ}$ increments, paralleling
the sequence in Figure~\ref{fig:lin-pol-out}. The top row is total
irradiance, indicating input polarization direction in the lower-left
corner of each panel; the middle row is the polarization component
parallel to the input polarization (indicated lower left in each panel);
the bottom row is the orthogonal polarization component (also indicated
lower left in each panel). At right is the experimental result corresponding
to the first column in the set of model results at left. Each frame
is $50\lambda/7D$ radians across. Intensities are normalized to the
same value in all frames.\label{fig:ffpat-lin-normal-az}}
\end{figure}

At normal incidence, the azimuthal orientation of the input polarization
impacts the output polarization state, as seen in
Figure~\ref{fig:lin-pol-out}.  Following the same CCR rotation sequence and
input polarization as was used in Figure~\ref{fig:lin-pol-out}, we produce
the far-field diffraction patterns in Figure~\ref{fig:ffpat-lin-normal-az}.
The polarization state of the central peak follows that of the input
polarization.  The total diffraction pattern rotates by $120^{\circ}$ as
the polarization rotates through $60^{\circ}$ in the opposite direction,
producing a net $180^\circ$ rotation of the pattern with respect to the
polarization state---just as the polarization ellipses did in
Figure~\ref{fig:lin-pol-out}.

\begin{figure}
\begin{center}\includegraphics[%
  height=2.5in,
  keepaspectratio]{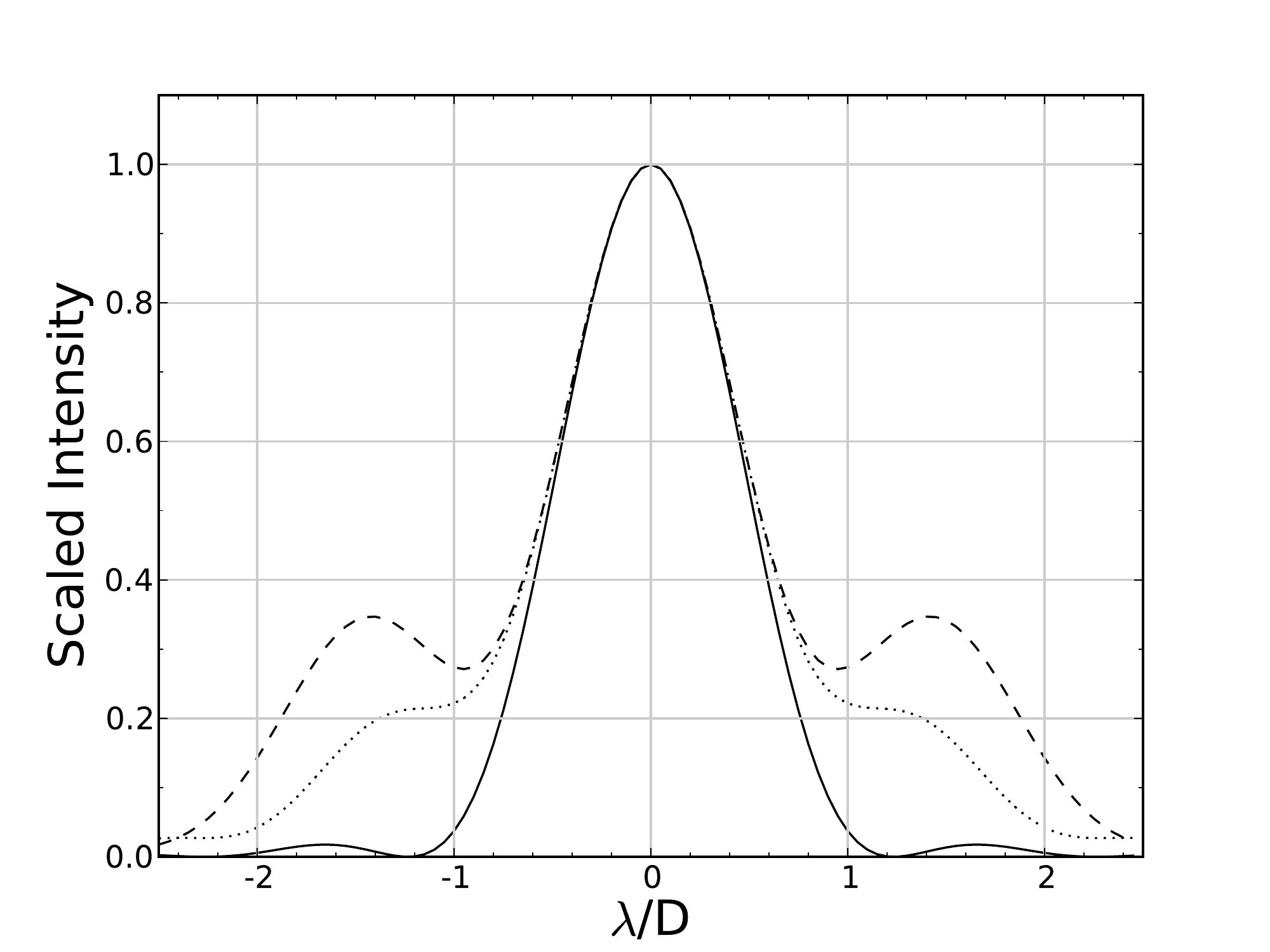}\end{center}

\caption{Orthogonal cuts (dashed and dotted) through the normal incidence
far-field diffraction pattern for the TIR CCR under linear input polarization,
showing the similarity of the central peak to the scaled Airy function (solid).
The cuts correspond to the upper left panel of Figure~\ref{fig:ffpat-lin-normal-az}.\label{fig:cuts-vs-airy}}
\end{figure}

Figure~\ref{fig:cuts-vs-airy} shows two profiles through the normal
incidence TIR CCR diffraction pattern compared to the scaled Airy pattern.
The two profiles correspond to orthogonal cuts through the center of the
pattern in the upper-left panel of Figure~\ref{fig:ffpat-lin-normal-az},
one of which passes through two outer lobes, and the other passing between
lobes.  The plot shows the symmetry of the central peak, and its similarity
to the Airy function over a considerable radius. In units of $\lambda/D$,
the TIR pattern departs from the Airy pattern by 1\% of full scale at a
radius of 0.30, by 5\% at 0.47--0.48, and by 10\% at 0.59--0.61, where
ranges refer to the two different profiles.  The functional form away from
the central peak is particularly relevant for satellite and lunar ranging
applications, where the tangential velocity of the target results in a
shift (velocity aberration) of the diffraction pattern at the position of
the transmitter, so that a co-located receiver samples the shoulder of the
diffraction pattern rather than its peak.  Lunar ranging to the 38~mm
diameter CCRs at 532~nm imposes a velocity aberration of 4--6~$\mu$rad,
which corresponds to about 0.29--0.43~$\lambda/D$.  The Airy function is
therefore still accurate to within 5\% in this regime, for normal incidence.

\begin{figure}
\begin{center}\includegraphics[%
  scale=0.5]{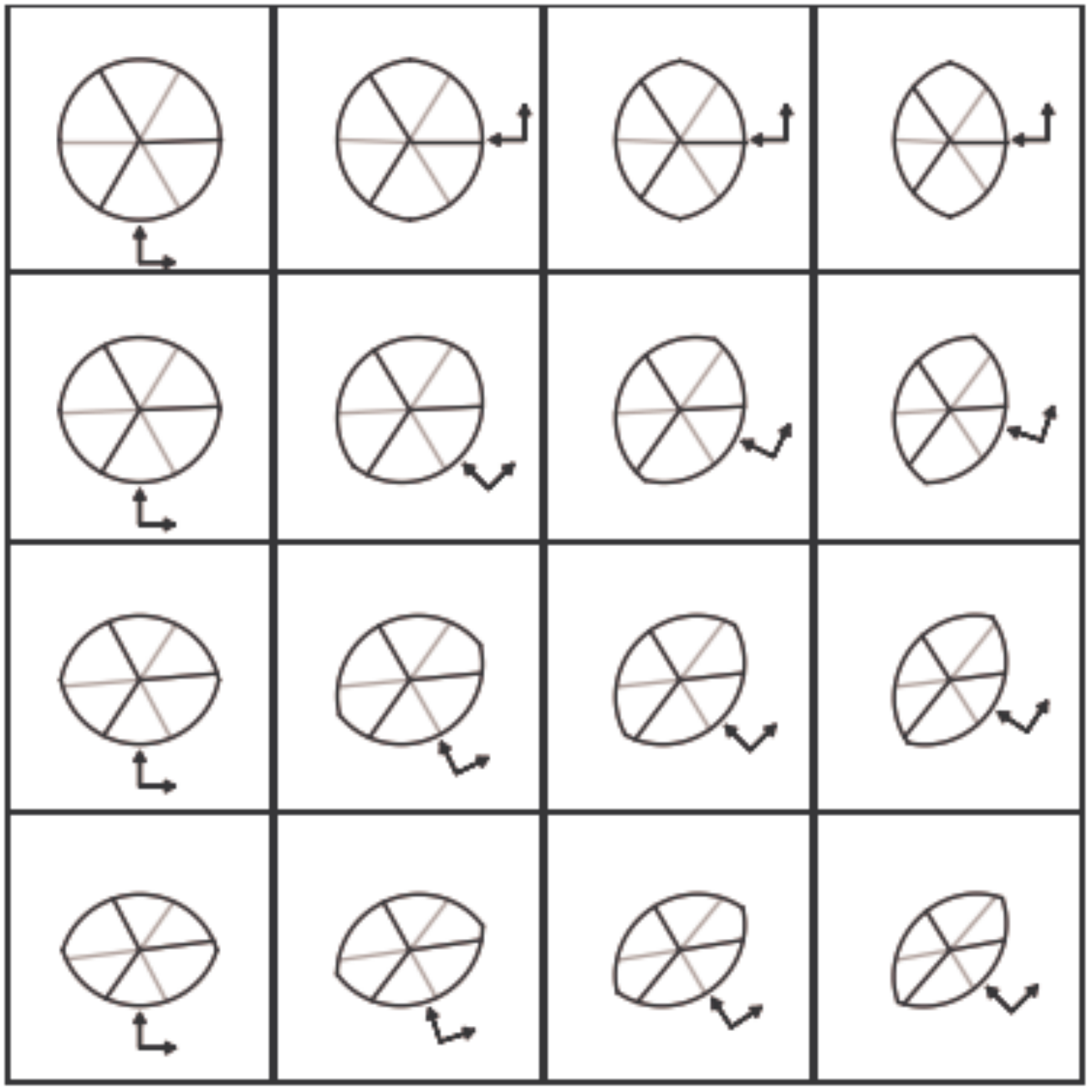}\end{center}

\caption{Orientation scheme and aperture shapes for the diffraction
patterns to follow. Normal incidence is at upper left, with each tile
representing a $5^{\circ}$ step along the positive-$x$ axis to the right
and along the negative-$y$ axis in the down direction. The
horizontal-vertical basis vectors ($\hat{\mathbf{s}}_0$ and
$\hat{\mathbf{p}}_0$) are placed at the azimuthal position of the observer,
vertical pointing toward the aperture. Black lines represent the refracted
appearance of real edges, while gray lines are the reflected
edges.\label{fig:aperture-grid}}

\end{figure}

As we move away from normal incidence, we may consider the effect
of azimuth and inclination angle on the patterns. We present results
in a $4\times4$ grid corresponding to off-axis positions on a $5^{\circ}$
pitch and in a Pythagorean arrangement. We place the normal incidence
case at the upper left, so that the fourth panel over in the top row
corresponds to an inclination angle of $15^{\circ}$ and an azimuth
of $A=0^{\circ}$, as defined in Section~\ref{sec:geometry}. This
corresponds to the distant observer placed in the positive-$x$ direction
in the global coordinate frame of Figure~\ref{fig:CCR-geom}, with
$y=0$. The second panel over in the bottom row has an inclination
angle of $\sqrt{15^{2}+5^{2}}\approx15.8^{\circ}$ and an azimuth
of $\arctan(-15/5)\approx-72^{\circ}$, putting the observer at a
positive-$x$ coordinate, with $y=-3x$. Figure~\ref{fig:aperture-grid}
shows the appearance of the effective apertures as seen by the observer
in these positions.

\begin{figure}
\includegraphics[%
  scale=0.33]{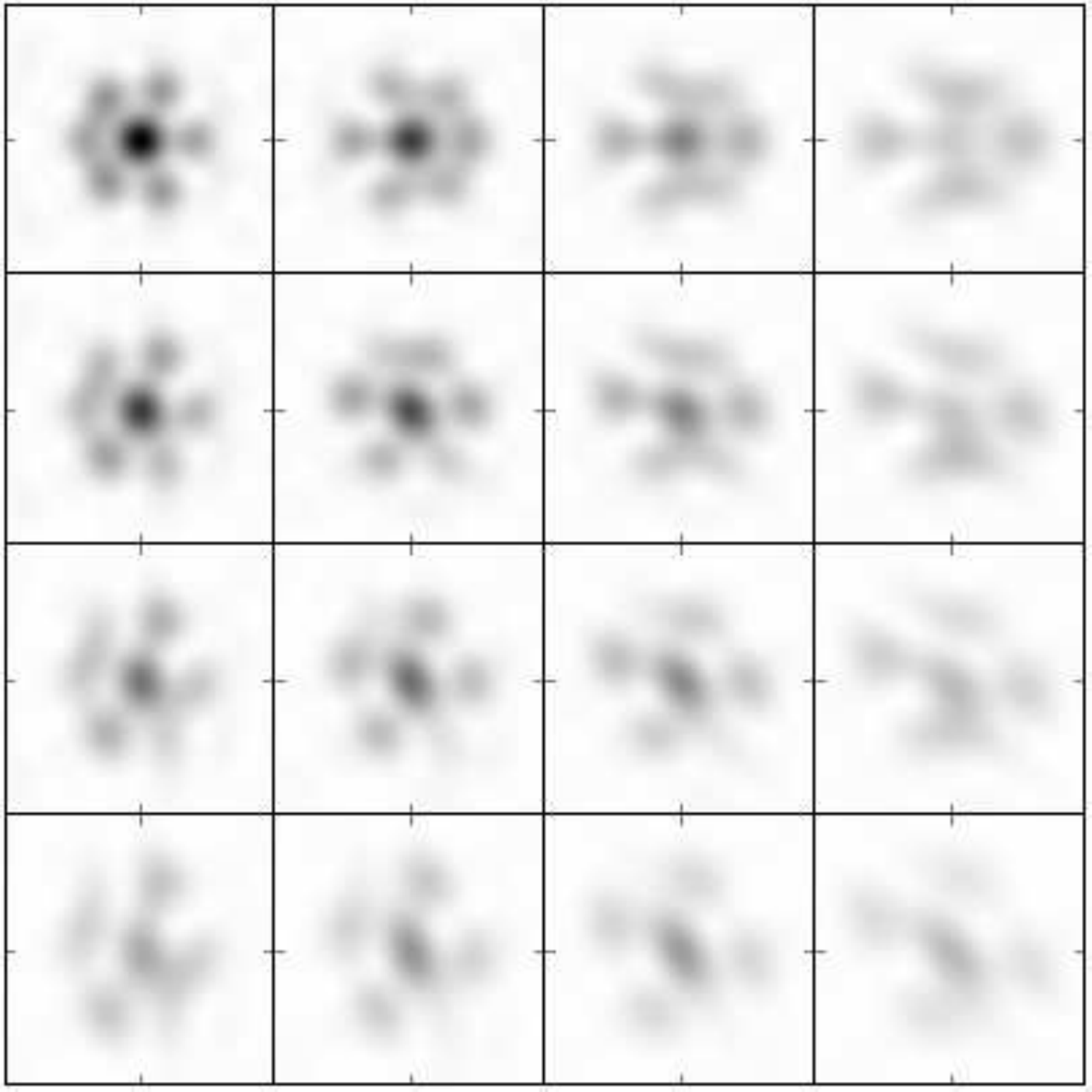}\hfill{}\includegraphics[%
  scale=0.33]{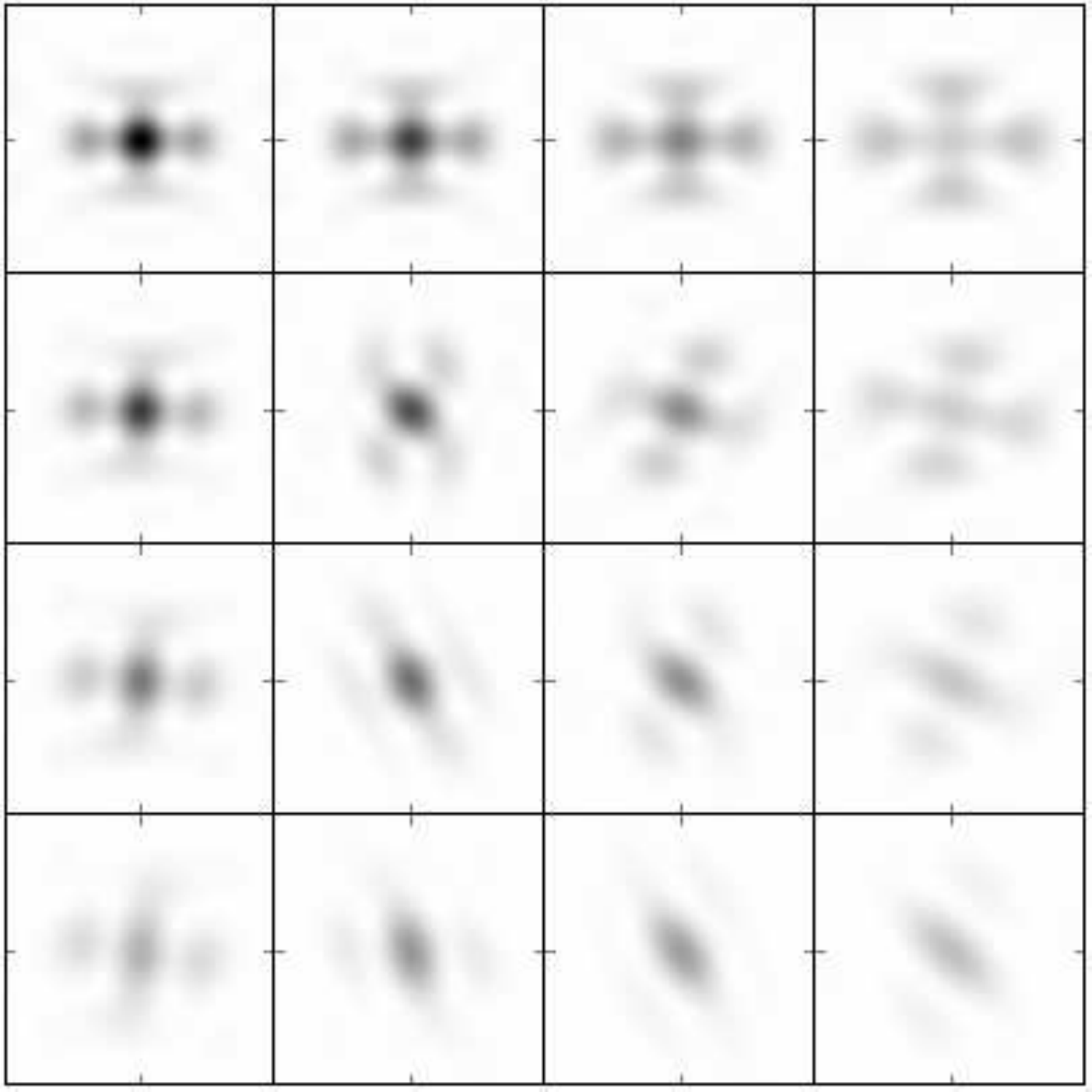}\hfill{}\includegraphics[%
  scale=0.33]{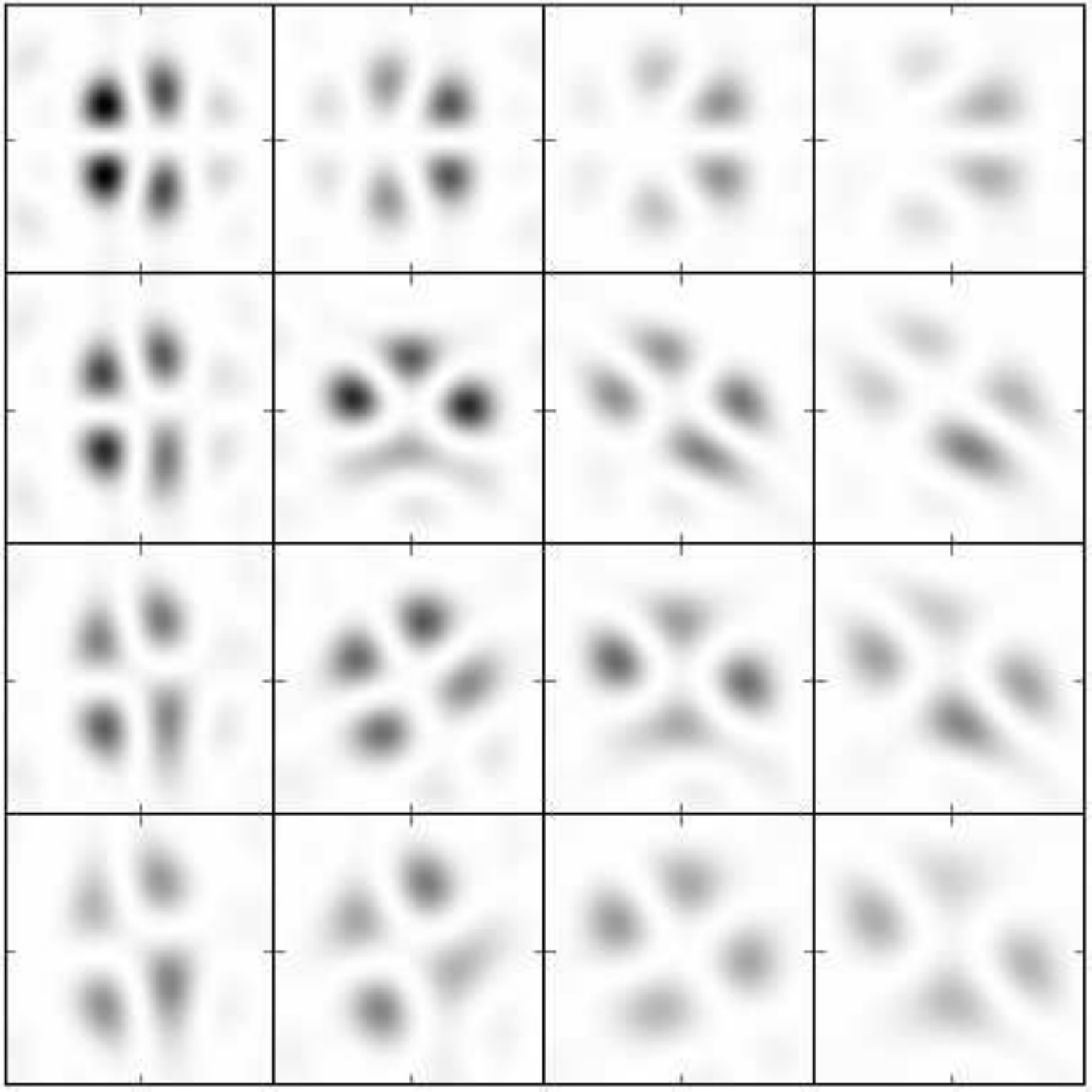}

\caption{Diffraction patterns for horizontal linear input polarization at
a range of viewing angles. In each grid, the upper left panel is at
normal incidence, following the orientation scheme depicted in Figure~\ref{fig:aperture-grid}.
At left is the total intensity, followed by the horizontal and vertical
polarization components. Intensities are normalized to a common maximum
within each of the three sets, but the intensity of the vertical polarization
panel has been scaled by a factor of 2.57 relative to the other two in
order to show details in these intrinsically dimmer patterns.\label{fig:ffdp-horz-grid}}
\end{figure}

For horizontal polarization input---which we define as perpendicular
to the plane of incidence---we get the patterns seen in Figure~\ref{fig:ffdp-horz-grid}.
For vertical input polarization, the patterns look the same, but with
a $180^{\circ}$ rotation of all frames and the middle panel corresponding to
vertical polarization output and the rightmost panel corresponding to
horizontal output.

\begin{figure}
\includegraphics[%
  scale=0.33]{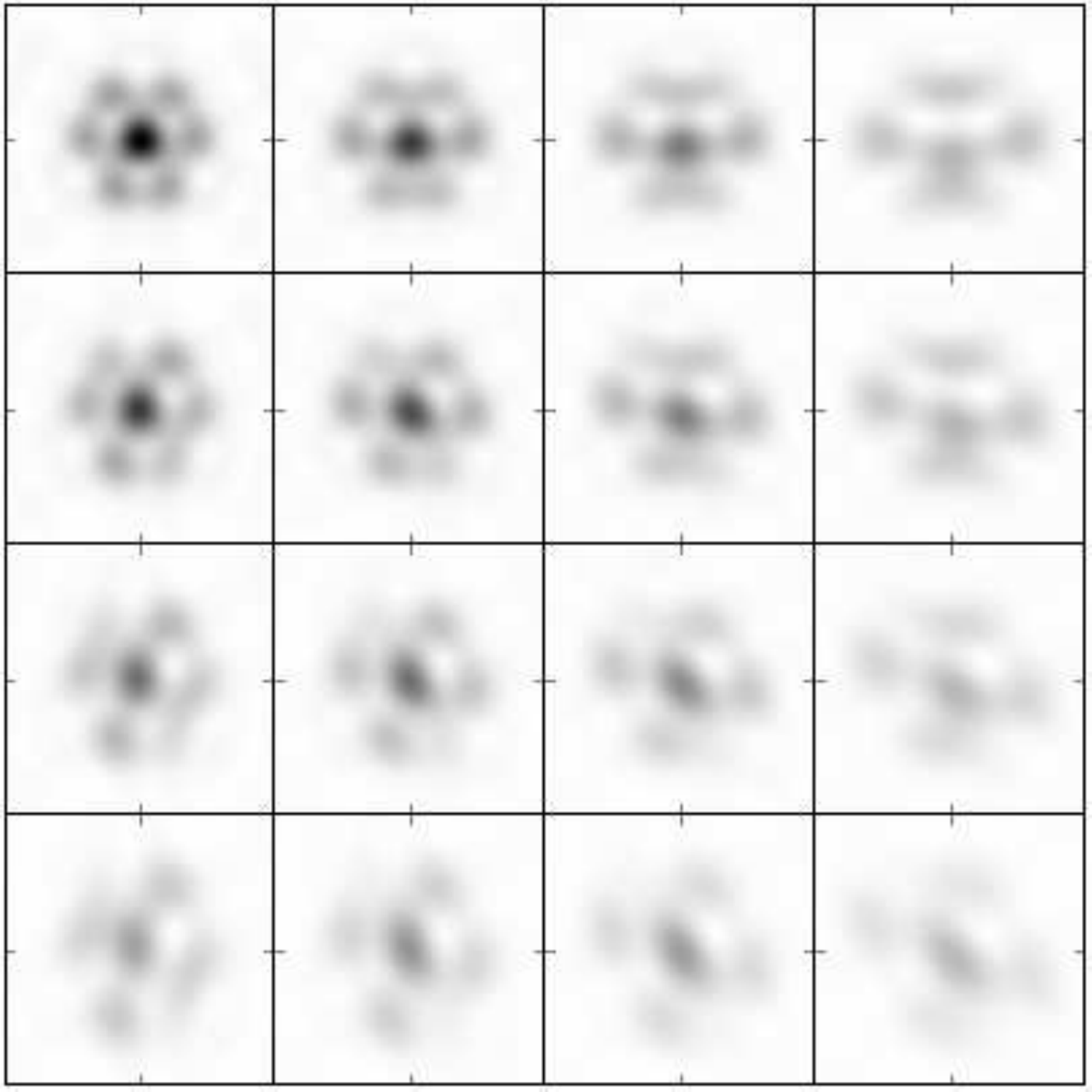}\hfill{}\includegraphics[%
  scale=0.33]{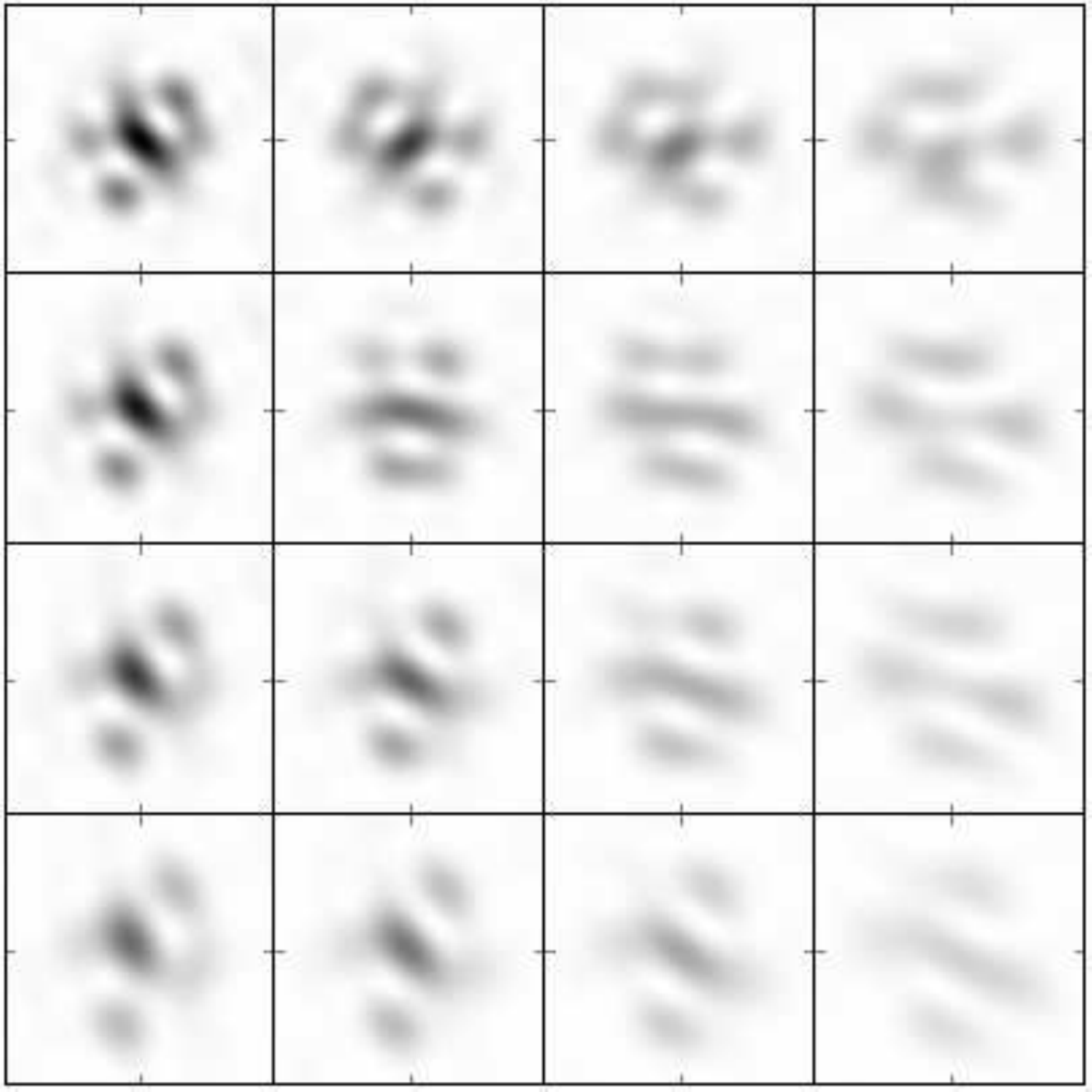}\hfill{}\includegraphics[%
  scale=0.33]{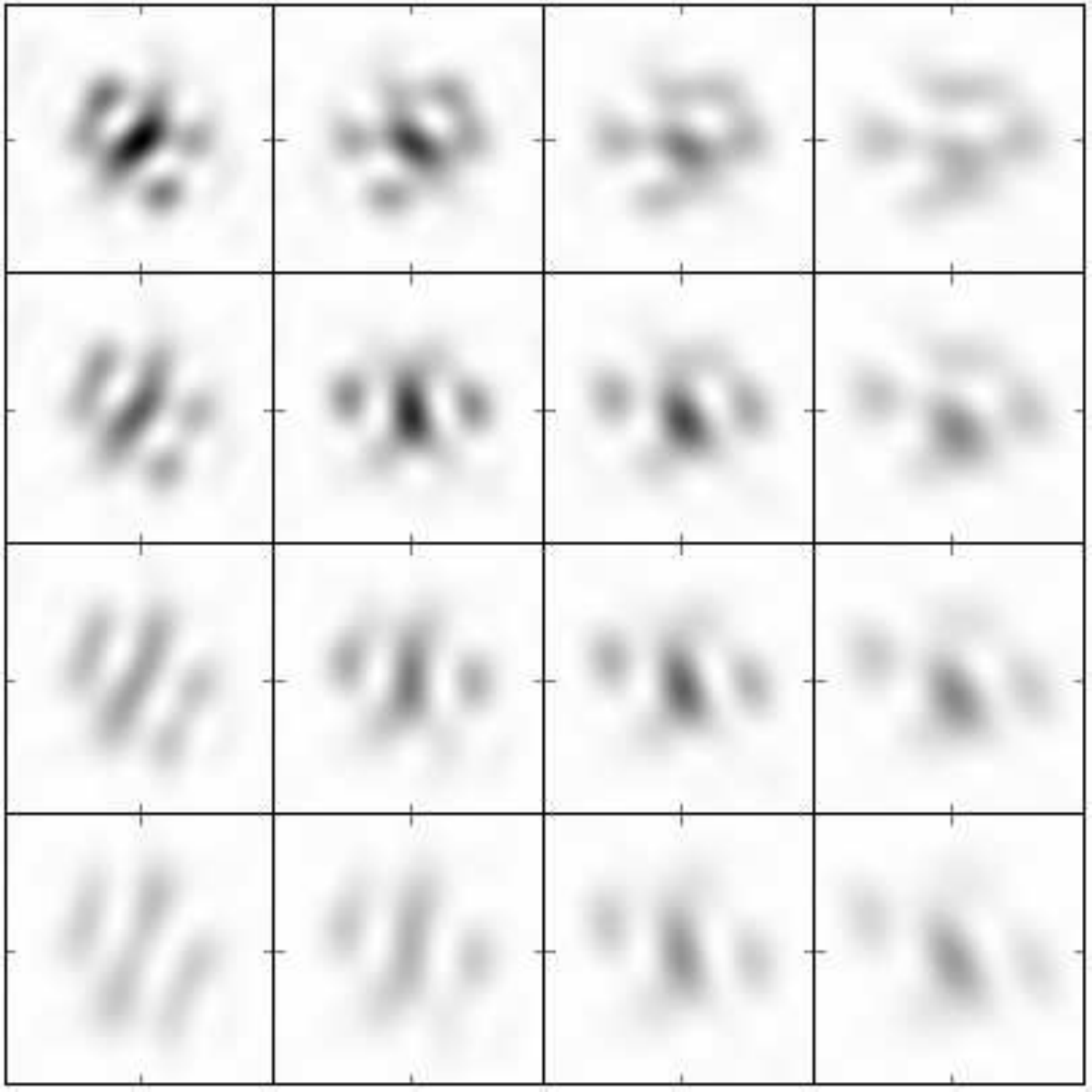}

\caption{Diffraction patterns for left-handed circular input polarization,
following the conventions of Figures~\ref{fig:aperture-grid} and
\ref{fig:ffdp-horz-grid}. The horizontal and vertical polarization
decomposition does not in this case break cleanly into distinct patterns as
was the case for linear input polarization, since the central spot in this
case is circularly polarized and thus shares equally in the two components.
\label{fig:ffdp-cw-grid}}
\end{figure}

The far-field diffraction patterns for left-handed circular input
polarization are shown in Figure~\ref{fig:ffdp-cw-grid}. The patterns
for right-handed circular polarization are the same except for a $180^{\circ}$
rotation of each frame and an exchange of horizontal and vertical
polarizations. The normal-incidence pattern has a three-fold symmetry lacking
in the linear polarization case, which stems from the complete
orientation-invariance of circular polarization, so that only the corner
cube asymmetries may imprint on the diffraction pattern.  Evidence for
symmetry is also clear in the polarization patterns of 
Figure~\ref{fig:circ-pol-out}.

\subsection{Laboratory Results\label{sub:exp-diffrac}}

We formed a linearly-polarized plane wave from a HeNe laser across a 25~mm
diameter, having a wavefront quality of approximately $\lambda/$4 as judged
visually by a shear plate. To achieve a uniform spatial intensity across
the aperture, we placed a $D=9.1$~mm circular aperture in front of the CCR,
concentric with and close to the front face (wavefront quality was better
over the smaller aperture).  The CCR used was a 25.4~mm diameter
high-precision fused silica corner cube. We also tested a flight spare CCR
from the Apollo retroreflector arrays, finding similar results---albeit with
increased scattered light and diffraction spikes owing to the
intentionally--ground edges where the rear CCR surfaces meet occupying a
significant fraction of the 9~mm aperture.

The beam passed through an uncoated fused silica wedge window having
$\lambda/10$ surface quality before striking the CCR at normal incidence.
The wedge window was tilted to reflect the returning beam away from the
optical axis by an angle of approximately $10^{\circ}$ for access to
imaging. A 339~mm focal length lens produced a far-field pattern onto a CCD
camera with 3.65~$\mu$m pixels. This results in 15.75 pixels spanning the
$2.44\lambda/D$ Airy diameter. Replacing the corner cube with a flat mirror
produced an Airy pattern having azimuthally uniform rings and approximately
84\% of the total flux within the first dark ring, as expected. The same
measure performed on the TIR pattern under horizontal polarization produced
$36.1\pm0.6$\%, in perfect agreement with the theoretical expectation.

\begin{figure}
\begin{center}\includegraphics[%
  width=5in,
  keepaspectratio]{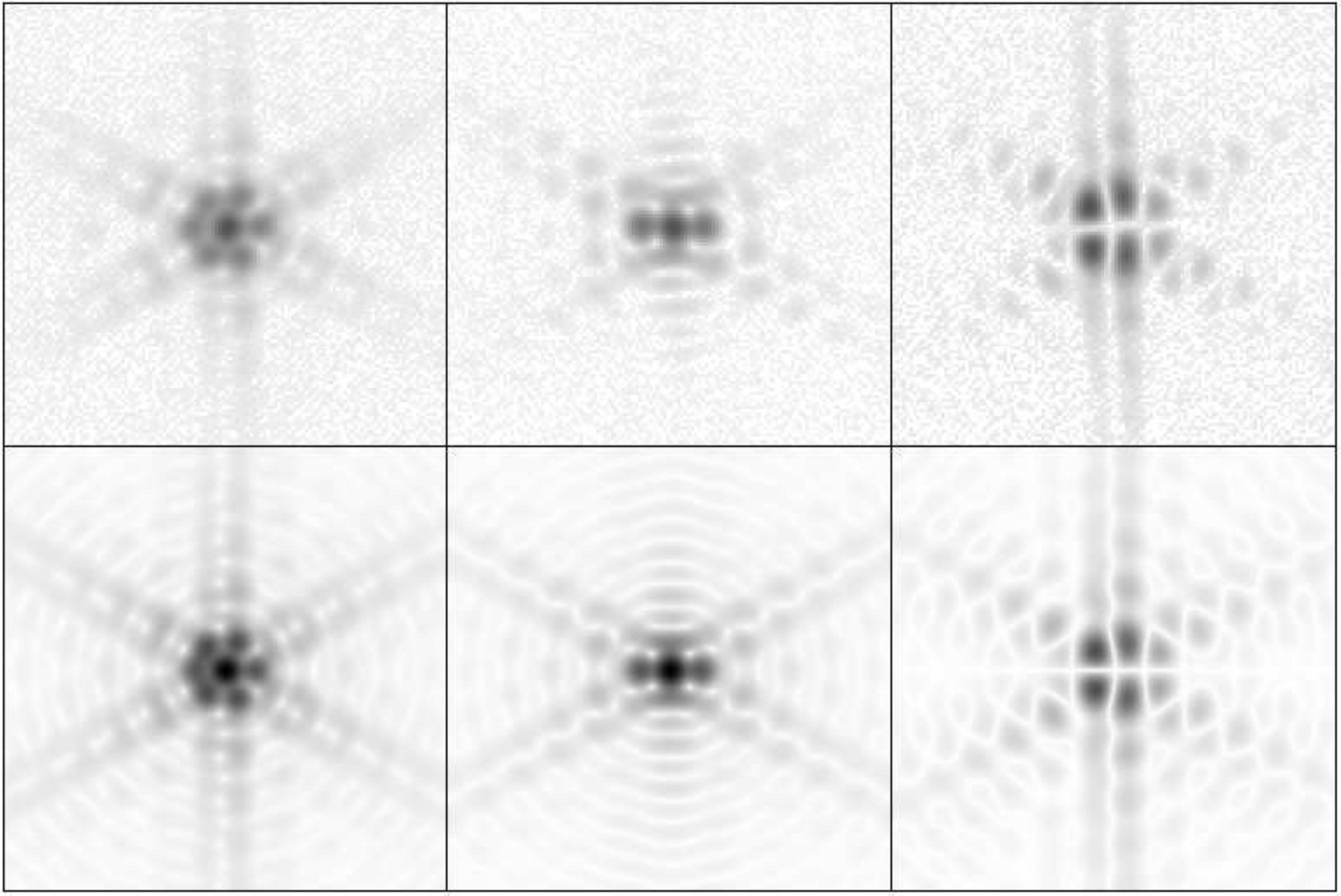}\end{center}

\caption{Comparison of experimental results (top) to the simulation results
(bottom), stretched to emphasize faint structure. At left is total
intensity given horizontal polarization input light, followed by horizontal
and vertical polarization output patterns.  Each frame is $64\lambda/3D$ across. \label{fig:experimental-diffrac}}
\end{figure}

The experimental diffraction pattern images in Figure~\ref{fig:experimental-diffrac}
have been rotated and reflected to place the experimental results
in the same frame established for the simulated patterns (i.e., transformations
followed the physical setup, and are not simply forced to match simulations).

\section{Conclusions}

The polarization states and resulting diffraction patterns from TIR CCRs are
non-trivial and generally require computational tools to assess. This
paper presents a comprehensive methodology for doing so, and provides
results against which independent analyses may be compared. The results
compare well against some---but not all---items available in the literature, and
laboratory measurements confirm the validity of the mathematical
development. The Python code that generated all simulation results
contained in this paper is available at 
\url{http://physics.ucsd.edu/~tmurphy/papers/ccr-sim/ccr-sim.html}.

This tool can provide a springboard from which one might analyze aberrations
from manufacturing imperfections, intentional offset angles of the
rear surfaces, thermally-induced refractive index gradients, aperture
masking or blockage, non-planar wavefront input, etc.  In a companion paper
\cite{ccr-thermal},
we explore the impact of thermal gradients on the diffraction patterns from
TIR corner cube prisms.

\section*{Acknowledgments}

We thank Jim Faller for loaning to us the Apollo flight spare corner cube.
We are also grateful to Ray Williamson for helping us sort out confusing
mis-information on wave plate orientations and tests to determine their
optical axes.  Part of this work was funded by the NASA Lunar Science
Institute as part of the LUNAR consortium (NNA09DB30A), and part by the
National Science Foundation (Grant PHY-0602507).

\end{document}